# IMPROVED SPATIAL MODULATION FOR HIGH SPECTRAL EFFICIENCY

Rajab M. Legnain, Roshdy H.M. Hafez[1] and Abdelgader M. Legnain[2]

[1]Department of Systems and Computer Eng., Carleton University, Ottawa, Canada
```
{rlegnain, hafez}@sce.carleton.ca
```
[2]Department of Electrical and Electronics Eng., Garyounis University, Benghazi, Libya.
```
legnain@ieee.com
```

*ABSTRACT*

*Spatial Modulation (SM) is a technique that can enhance the capacity of MIMO schemes by exploiting the index of transmit antenna to convey information bits. In this paper, we describe this technique, and present a new MIMO transmission scheme that combines SM and spatial multiplexing. In the basic form of SM, only one out of $M_T$ available antennas is selected for transmission in any given symbol interval. We propose to use more than one antenna to transmit several symbols simultaneously. This would increase the spectral efficiency. At the receiver, an optimal detector is employed to jointly estimate the transmitted symbols as well as the index of the active transmit antennas. In this paper we evaluate the performance of this scheme in an uncorrelated Rayleigh fading channel. The simulations results show that the proposed scheme outperforms the optimal SM and V-BLAST (Vertical Bell Laboratories Layered space-time at high signal-to-noise ratio (SNR). For example, if we seek a spectral efficiency of 8 bits/s/Hz at bit error rate (BER) of $10^{-5}$, the proposed scheme provides 5dB and 7dB improvements over SM and V-BLAST, respectively.*

## 1. INTRODUCTION

Wireless communication systems using MIMO (Multiple Input Multiple Output) have been shown to achieve significantly higher spectral efficiencies than conventional single-antenna systems. In [1], this performance improvement was demonstrated using the Vertical-Bell Laboratories Layered Space-Time (V-BLAST) system.

In [2][3], the spatial modulation (SM) technique was introduced. Instead of the normal two-dimensional modulation (e.g. QAM), the SM introduces a third dimension which is the index of the antenna where the symbol is emitted from. In the basic form of SM, the transmitter has access to $M_T$ antennas, but only one out of the $M_T$ antennas is used to transmit in any given symbol interval. The receiver must determine which of the $M_T$ antennas was selected for transmission. The choice of one out of $M_T$ antennas conveys $\log_2(M_T)$ bits of information. At the receiver, iterative-maximum ratio combining (i-MRC) is used to estimate both the transmitted symbol and the index of the active antenna. This technique achieves comparable performance with V-BLAST, but with significantly lower complexity at the receiver [2].

In [4] Jeganathan et al. proposed an optimal detector for SM, which showed significant improvement over V-BLAST and conventional SM (SM with i-MRC detector) with reasonable increase in receiver complexity. In [5] Younis et al. proposed the Sphere decoder (SD) for SM to reduce the receiver complexity. It was shown that SM with SD can achieve comparable performance to the optimal SM decoder with lower complexity.

## 2. RELATED WORK

In [6] and [7] Jeganathan et al. presented a new modulation scheme based on SM, called generalized space shift keying (GSSK) and space shift keying (SSK), respectively. In SSK, the information bits are conveyed using the antenna index, and in GSSK, the information bits are conveyed using combinations of active antenna indexes.

Generalised SM was proposed in [8] and [9] which extended the concept of SM. In these schemes, at each time interval one symbol is transmitted by a combination of active antennas.

In [10] Başar et al. used Space-Time Block coding (STBC) for SM. The STBC-SM applies the transmit diversity of the STBC to the SM scheme. Their simulation results showed that the STBC-SM provides better performance compared to SM and can offer 3-5 dB improvement in bit error rate performance over SM and V-BLAST (depending on spectral efficiency).

In this paper, we propose a new MIMO transmission scheme based on SM. Unlike the SM which uses only one out of $M_T$ antennas to transmit a symbol in any given symbol interval [2][3], the proposed scheme uses $M_A$ antennas to transmit $M_A$ symbols simultaneously, where $M_A < M_T$. The proposed scheme groups a sequence of independent random bits into blocks, in which each block contains $\log_2(M_T M^{M_A})$ bits, ($M$ is modulation order). The first $\log_2(M_T)$ bits are used to select $M_A$ transmit antennas, with indexes $i, i+1, (i+M_A-1)$, where $i$ is the antenna index. Then, the last $\log_2(M^{M_A})$ bits are transmitted on the $M_A$ antennas after being modulated using a conventional modulation scheme (e.g., $M$-PSK, $M$-QAM, etc.). For example, consider $M_T = 4$, $M_A = 2$, and $M = 2$, then four bits can be transmitted simultaneously. Suppose a block of four bits to be transmitted as, say [0 1 1 0], then, the active antenna will be antenna indexes 2 and 3, and the transmit symbols will be 1 and -1 on antenna 2 and 3, respectively. Table 1 illustrates the mapping of the proposed scheme. Note that, when $M_A = 1$ the scheme becomes conventional SM.

Table 1. Proposed scheme mapping: $M_A = 2$, $M_T = 4$, M = 2.

| Block Input | Active antenna | Transmit symbol vector, x |
|---|---|---|
| 0000 | 1,2 | [-1 -1 0 0 ] |
| 0001 | 1,2 | [-1 1 0 0 ] |
| 0010 | 1,2 | [ 1 -1 0 0] |
| 0011 | 1,2 | [ 1 1 0 0 ] |
| 0100 | 2,3 | [0 -1 -1 0] |
| 0101 | 2,3 | [ 0 -1 1 0] |
| 0110 | 2,3 | [ 0 1 -1 0] |
| 0111 | 2,3 | [ 0 1 1 0] |
| 1000 | 3,4 | [0 0 -1 -1] |
| 1001 | 3,4 | [ 0 0 -1 1] |
| 1010 | 3,4 | [0 0 1 -1 ] |
| 1011 | 3,4 | [ 0 0 1 1 ] |
| 1100 | 4,1 | [-1 0 0 -1] |
| 1101 | 4,1 | [1 0 0 -1 ] |
| 1110 | 4,1 | [-1 0 0 1 ] |
| 1111 | 4,1 | [ 1 0 0 1 ] |

The rest of the paper is organized as follows: In Section 3 we describe the mapper and the detector of the new MIMO transmission scheme. Simulation results and conclusion are provided in Section 4 and 5, respectively.

Throughout the paper, the following notations are used. Bold lowercase and bold uppercase letters denote vectors and matrices, respectively. We use $[.]^T$, $Tr[.]$, $[.]^*$, and $[.]^H$ to denote transpose, trace, conjugate and Hermitian of a matrix or a vector, respectively. Furthermore, we use $\|.\|_F$ to denote Frobenius norm of a matrix or a vector, and $E[.]$ to denote the expectation.

## 3. SYSTEM MODEL

We consider a MIMO system where $M_T$ is the total number of transmit antennas and $M_R$ is the total number of receive antennas. During a transmission period, the number of active antennas is $M_A$, ($M_A < M_T$). All channel elements are assumed to be mutually uncorrelated flat fading channels. The system model of the proposed scheme is shown in Fig. 1. In the Figure, **b** is a sequence of independent random bits to be transmitted. The new scheme groups the incoming bits into blocks of $\log_2(M_T M^{M_A})$ bits. Each block is mapped into a vector, which is then transmitted over MIMO channel, (e.g. see table 1)

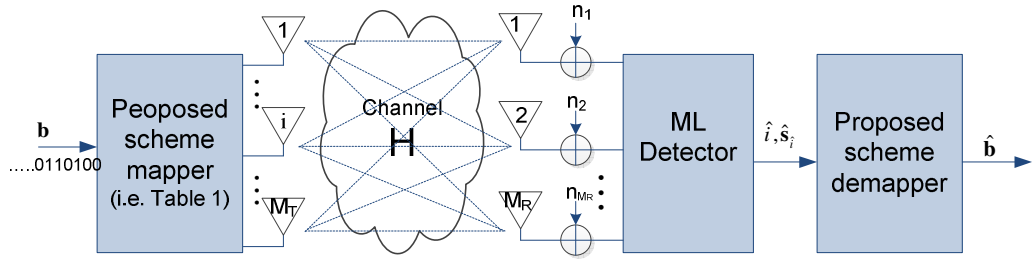

Figure 1. The proposed scheme system model

The $M_A \times 1$ transmitted signal vector is given by

$$\mathbf{x} = \mathbf{s}^{\Xi_i}, \qquad (1)$$

where $\Xi$ denotes the Circular shift operation, for example, if $\mathbf{a} = [a_1, a_2, ..., a_M]^T$ then $\mathbf{a}^{\Xi_2} = [a_M, a_1, ..., a_{(M-1+2)}]^T$. $\mathbf{s} = [s_1, s_2, ..., s_{M_A}, 0, ..., 0_{(M_T - M_A)}]^T$, $s_1, s_2, ..., s_{M_A}$ are the transmitted symbols which are selected from an $M$-ary signal constellation. The covariance matrix of **s**, $\mathbf{R}_{ss} = E[\mathbf{s}\mathbf{s}^H]$, must satisfy the power constraint, $Tr[\mathbf{R}_{ss}] = E_s$. In other words, The average transmit energy per symbol is $\frac{E_s}{M_A}$, where $E_s$ is the average transmit energy. Then, the proposed scheme can transmit $\log_2(M_T M^{M_A})$ bits simultaneously.

At the receiver, the received sample vector on the receive antennas can be expressed as

$$\mathbf{y} = \mathbf{H}\mathbf{x} + \mathbf{n} \qquad (2)$$

where $\mathbf{y} = [y_1, y_2, ..., y_{M_R}]^T$ is the $M_R \times 1$ received sample vector, and $\mathbf{n} = [n_1, n_2, ..., n_{M_R}]^T$ is the $M_R \times 1$ additive noise vector, in which each element is assumed to be an independent and identically distributed (iid) zero mean complex Gaussian random variable with variance $\sigma_N^2$. **H** is the channel matrix between transmit antennas and receive antennas, and it is given by

$$\mathbf{H} = \begin{bmatrix} h_{1,1} & h_{1,2} & \cdots & h_{1,M_T} \\ h_{2,1} & h_{2,2} & \vdots & \vdots \\ \vdots & \vdots & \ddots & \vdots \\ h_{M_R,1} & h_{M_R,2} & \cdots & h_{M_R,M_T} \end{bmatrix}, \quad (3)$$

where $h_{j,i}$ is the complex fading coefficient between the $i^{\text{th}}$ transmit antenna and the $j^{\text{th}}$ receive antenna. $h_{j,i}$ is assumed to be iid complex zero mean Gaussian with variance one.

The receiver uses maximum likelihood detector to estimate the index, $\hat{i}$, and the transmitted symbol vector, $\hat{\mathbf{s}}$. The ML detector estimates the index $i$, and the transmitted symbol vector, **s**, as [11]

$$[\hat{i}, \hat{\mathbf{s}}] = \arg\max_{i,\mathbf{s}} \Pr(\mathbf{y} \mid \mathbf{H}, \mathbf{s}^{\Xi_i}) \quad (4)$$

$$= \arg\min_{i,\mathbf{s}} \left( \|\mathbf{y} - \mathbf{H}, \mathbf{s}^{\Xi_i}\|_F^2 \right)$$

where

$$\Pr(\mathbf{y} \mid \mathbf{H}, \mathbf{s}^{\Xi_i}) = \frac{1}{\pi^{M_R} \sigma_N^{2M_R}} \exp\left( \frac{\|\mathbf{y} - \mathbf{H}, \mathbf{s}^{\Xi_i}\|_F^2}{\sigma_N^2} \right) \quad (5)$$

is the conditional probability density function (PDF) of **y** given **H** and $\mathbf{s}^{\Xi_i}$. Equation (4) estimates both the index and the transmitted symbols jointly by searching over all combination of the index $i$ and the symbol vector *s*.

## 4. SIMULATION RESULTS

In this section, we provide simulation results for the proposed MIMO transmission scheme and compare it with the results of optimal SM and V-BLAST. The V-BLAST system uses minimum mean square error ordered successive interference cancellation (MMSE-OSIC) detection [12]. Monte Carlo simulations are used to evaluate bit error rate (BER) performance of the proposed scheme, SM and V-BLAST for different spectral efficiencies ($\eta$), number of transmit antennas ($M_T$) and number of active antennas ($M_A$). We assume uncorrelated flat Rayleigh fading channel. *M*-QAM modulation with Gray mapping is used in the simulation.

Figure 2 shows the BER performance for 6 bits/s/Hz of $4 \times 4$ 4-QAM new scheme with $M_A=2$, $4 \times 4$ 16-QAM, SM and $3 \times 4$ 4-QAM V-BLAST. From Figure 2, the new scheme outperforms the optimal SM and V-BLAST. At BER of $10^{-5}$, the new scheme provides SNR gain of about 2.2 dB over SM and V-BLAST.

Figure. 3 shows the BER performance for 8 bits/s/Hz of $4 \times 4$ 4-QAM new scheme with $M_A=3$, $4 \times 4$ 64-QAM, SM and $4 \times 4$ 4-QAM V-BLAST. From the figure, the new scheme provides SNR gains of 5 dB and 7 dB over optimal SM and V-BLAST at BER of $10^{-5}$, respectively.

Figure. 4 shows the BER performance for 12 bits/s/Hz of $8\times 4$ 8-QAM with $M_A$=3, and $4\times 4$ 32-QAM with $M_A$=2 for the new scheme, $8\times 4$ 512-QAM for SM and $4\times 4$ 8-QAM V-BLAST. It can be seen that, the new scheme provides huge performance improvement over optimal SM and V-BLAST. At $10^{-3}$, the $8\times 4$ new scheme provides about 7 dB and 10 dB SNR gains over optimal SM V-BLAST, respectively.

From Figures 2, 3 and 4, we conclude that the proposed scheme has better BER performance than V-BLAST, because it provides a full diversity order and it uses the index of antenna to convey information which leads to lower the modulation order. Also the proposed scheme outperforms the optimal SM, due to the fact that the modulation order used in the proposed scheme is lower than that used in optimal SM.

## 5. CONCLUSIONS

In this paper, we proposed a new MIMO transmission scheme to improve the spectral efficiency. In the scheme, we combine SM with spatial multiplexing. This new proposed SM scheme uses several antennas to transmit different symbols at the same time slot, where the active antennas are subset of a larger set of antennas. By computer simulation, BER performance for the proposed scheme was evaluated for uncorrelated Rayleigh fading channel and was compared to optimal SM and V-BLAST.

The simulation results show that the new MIMO transmission scheme outperforms optimal SM and V-BLAST at high SNR. Furthermore, the performance improvement of the new scheme over optimal SM and V-BLAST increases as the transmission rate increases, which makes it a potential candidate for high data rate transmission systems e.g., WiMAX and LTE-Advanced.

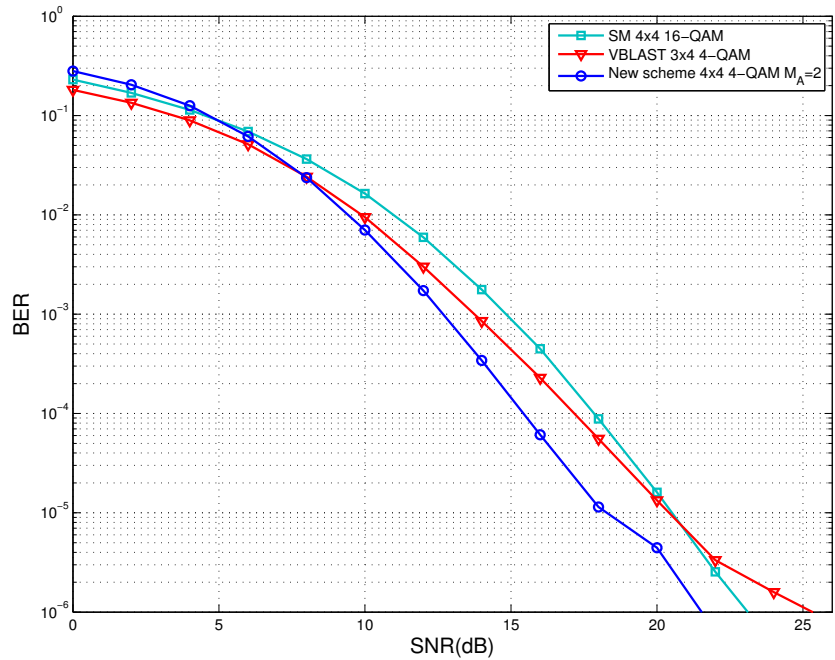

Figure 2. BER performance for 6 bits/s/Hz.

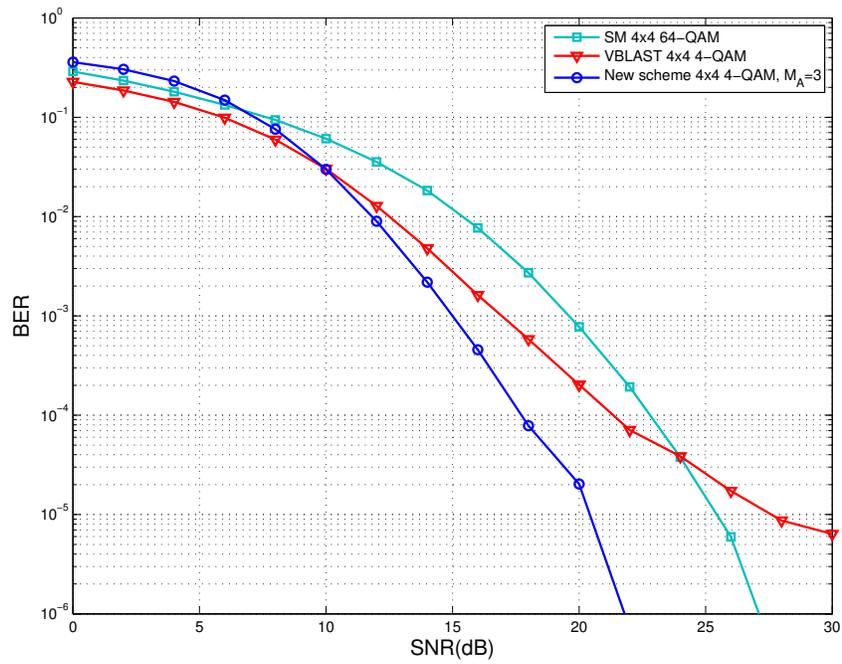

Figure 3. BER performance for 8 bits/s/Hz.

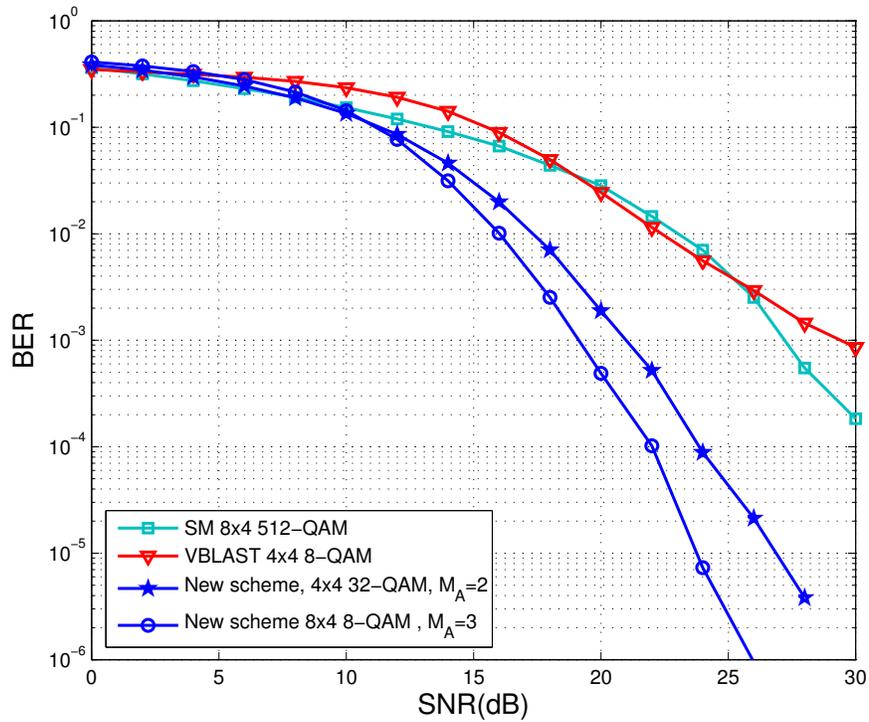

Figure 4. BER performance for 12bits/s/Hz.

**Authors**

**Rajab M. Legnain** is a Ph.D student at Systems and Computer Engineering, Carleton University, Ottawa, Canada. He received a B.Sc. and M.A.Sc in Electrical and Electronics Engineering both from University of Garyounis, Benghazi, Libya, in 2000 and 2004 respectively. He had been a faculty member of University of Garyounis at Electrical and Electronics Department from 2006 to 2008. His current research focuses on smart antennas and relay networks.

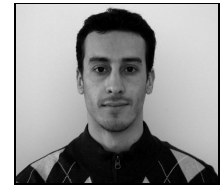

**Roshdy H.M. Hafez** obtained the Ph.D. in Electrical Engineering, from Carleton University, Ottawa, Canada. He joined the Department of Systems and Computer Engineering, Carleton University as an assistant professor, and he is now a full professor. Dr. Hafez has many years' experience in the areas of Wireless communications, RF and spectrum engineering. He has lectured extensively in wireless and related areas. His current research focuses on broadband wireless networking, 3G/4G/LTE, wireless over fiber and sensor networks.

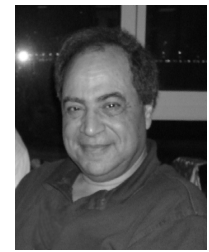